\documentstyle[12pt,epsf]{article}

\advance\voffset by -1.5cm
\advance\hoffset by -1.5cm
\def\be{\begin{equation}}
\def\ee{\end{equation}}

\def\pmb#1{\setbox0=\hbox{#1}
 \kern-.025em\copy0\kern-\wd0
 \kern.05em\copy0\kern-\wd0
 \kern-.025em\raise.0433em\box0 }

\def\3{\ss}
\def\sq{\hbox{\rlap{$\sqcap$}$\sqcup$}}
\def\qed{\ifmmode\sq\else{\unskip\nobreak\hfil
\penalty50\hskip1em\null\nobreak\hfil\sq
\parfillskip=0pt\finalhyphendemerits=0\endgraf}\fi}

\def\bbbz {{\sf Z\!\!Z}}
\def\bbbr {{\rm I\!R}}

\def\Tr{{\rm Tr}}

\def\bg{{\bf g}}
\def\bq{{\bf q}}
\def\bn{{\bf n}}

\def\bbeta{\mbox{\boldmath{$\beta$}}}
\def\bgamma{\mbox{\boldmath{$\gamma$}}}

\def\ss{\bf S}

\def\C{{\cal C}}

\begin{document}

\thispagestyle{empty}
\def\thefootnote{\fnsymbol{footnote}}
\begin{flushright}
  hep-th/9712211\\
  HUTP-97/A104
 \end{flushright}
\vskip 0.5cm

\begin{center}\LARGE
{\bf Three-Pronged Strings and 1/4 BPS States in N=4 
Super-Yang-Mills Theory}
\end{center}
\vskip 1.0cm
\begin{center}
    {\large  Oren Bergman\footnote{E-mail  address: {\tt
bergman@string.harvard.edu}} }

\vskip 0.5 cm
{\it Lyman Laboratory of Physics\\
Harvard University\\
Cambridge, MA 02138}
\end{center}

\vskip 1.5cm

\begin{center}
December 1997
\end{center}

\vskip 1.5cm

\begin{abstract}
We provide an explicit construction of 1/4 BPS states 
in four-dimensional $N=4$ Super-Yang-Mills theory with a
gauge group $SU(3)$. These states correspond to three-pronged
strings connecting three D3-branes.
We also find curves of marginal stability
in the moduli space of the theory, at which the above states
can decay into two 1/2 BPS states.
\end{abstract}

\vskip 1.5cm 
\begin{center}
\end{center}

\vfill
\setcounter{footnote}{0}
\def\thefootnote{\arabic{footnote}}
\newpage

\renewcommand{\theequation}{\thesection.\arabic{equation}}

\section{Introduction}
\setcounter{equation}{0}

\noindent $N=4$ super-Yang-Mills theory is the simplest
setting for the Olive-Montonen duality conjecture \cite{om}.
This theory is in fact believed to possess an exact $SL(2,\bbbz)$
symmetry \cite{sen1}, which includes an element corresponding
to the original electric-magnetic duality of \cite{om}.
This symmetry implies the existence of an infinite set of stable
dyon states, obtained by acting with the duality group
on the perturbative $W$-boson states. Stability of these
states is guaranteed by the fact that they preserve 1/2
of the underlying supersymmetry, and therefore transform
in short multiplets of the superalgebra \cite{ow,osborn}.
Existence is a more difficult question, and relies on identifying
harmonic forms on the moduli space of monopoles. 
In the simplest case of $SU(2)$ with two monopoles this
form was found by Sen \cite{sen1}. There are also indications
that such forms exist as well for more than two monopoles 
\cite{ss,porrati}. For higher rank gauge groups the strategy
is to embed the known $SU(2)$ solutions using the simple
roots of the group \cite{gl,gauntlett}.

D-branes \cite{poltasi} have allowed us to approach this problem from a new
direction. The low-energy world-volume theory of $n$ parallel
D$p$-branes is $p+1$-dimensional super-Yang-Mills theory
with sixteen supersymmetries, and a gauge group $U(n)$, 
which is spontaneously broken to $U(1)^n$ if none of the D-branes
coincide.
In particular, for D3-branes in 
type IIB string theory we get precisely four-dimensional $N=4$ SYM
\cite{tseytlin,gg}. 
In this picture the $W$-bosons appear as fundamental strings
between different D3-branes. Monopoles and dyons correspond
to D-strings and bound states of D-strings and fundamental strings,
respectively, between different D3-branes. These bound states are
required by $SL(2,\bbbz)$ duality of type IIB string
theory, and in fact shown to exist in \cite{wittenboundstates}.

Recently, a new kind of object has entered the game in type IIB
string theory, the so-called ``three-string junction" or 
``three-pronged string". The basic configuration corresponds
simply to a fundamental string ending on a D-string. The subtle
observation made by Schwarz \cite{schwarz} is that charge 
conservation requires that on one side the D-string turn into
a $(1,1)$ string. Other three-pronged strings can be
obtained from the basic one by acting with $SL(2,\bbbz)$.
This will ensure that the $(p,q)$ charges are always conserved.
More generally, three-pronged strings are believed to exist
for {\em any} three strings with relatively prime charges $(p_i,q_i)$,
under the condition that the total $p$ charge and total $q$ charge
vanish.
Furthermore, these objects were argued to be BPS saturated
if the angles between the three strings are adjusted properly
\cite{dm,sen2,ry,kl,mo}. 
Multi-pronged strings have been used to explain
the appearance of exceptional gauge symmetry enhancement in
eight dimensions \cite{gz}, and to construct certain BPS
states in five-dimensional field theories \cite{ahk}.

Since D3-branes are self-dual, {\it i.e.} invariant under $SL(2,\bbbz)$,
any $(p,q)$ string can end on them. It is therefore fair to assume that
they are allowed boundaries for three-pronged strings as well. 
The obvious question is what state such a configuration corresponds
to in the world-volume theory. 
Since one needs at least three D3-branes to support a three-pronged
string, we only expect these states to arise for gauge groups
$SU(3)$ and higher.

We will argue that these are the 1/4 BPS states of $N=4$ SYM,
by showing that their supersymmetry transformation properties
agree, as well as their gauge charges and masses. We will further
argue that these states exist only in certain regions of moduli space,
and derive the curves of marginal stability explicitly. On these curves
the 1/4 BPS states are unstable to decay into two states. This corresponds
to a point where one of the prongs degenerates, and the two remaining
prongs separate along the common D3-brane into two open strings.

The paper is organized as follows. In section~2 we review the
BPS states of $N=4$ SYM and their representation in terms of 
strings between D3-branes. In section~3 we discuss three-pronged
strings in general, and in the presence of D-branes. In section~4
we make the connection between three-pronged strings connecting
D3-branes and 1/4 BPS states of $N=4$ SYM. We also derive
the curves of marginal stability for these states, and explain
the decay process in terms of the string picture. In section~5
we present a summary of the results, and offer some directions
for future work.

\section{BPS States in N=4 SYM}
\setcounter{equation}{0}
\subsection{BPS states}

\noindent Consider four-dimensional $N=4$ super-Yang-Mills theory
with a simple gauge group $G$. This theory can be obtained
as the dimensional reduction of ten-dimensional $N=1$ super-Yang-Mills
theory on $T^6$ \cite{osborn}. The ten-dimensional Lorentz group reduces
to $SO(3,1)\times SO(6)$, where $SO(6)$ becomes a global R-symmetry.
The vector supermultiplet consists of a gauge field $A_\mu$
($\mu=0,\ldots,3$), six scalar fields 
$\phi^I$ ($I=4,\ldots,9$), transforming in the ${\bf 6}$ of $SO(6)$, and
four Weyl fermions transforming in the ${\bf 4}$ of $Spin(6)$,
all valued in the adjoint representation of $G$. The bosonic part of the action 
is given by
\be
  S = -{1\over 16\pi} \mbox{Im}
                  \int \tau\Tr(F\wedge F + iF\wedge *F)
        - {1\over 2e^2}\int \Tr\Big( |D\phi^I|^2 + 
     \sum_{I<J}[\phi^I,\phi^J]^2 \Big)\; ,
\label{action}
\ee
where $\tau=i/e^2 + \theta_{YM}/{2\pi}$. We shall assume that
$\theta_{YM}=0$.
The $N=4$ supersymmetry algebra admits two central charges, given by
\cite{osborn,fh}
\be
  Z^2_{\pm} =  |Q^I_E|^2 + |Q^I_M|^2 \pm 2|Q_E^I| |Q_M^I| \sin{\gamma} 
   \qquad (\gamma \geq 0) \; ,
\ee
where 
\be 
 Q_E^I = \int d{\bf S} \cdot \Tr({\bf E}\phi^I) \qquad , \qquad
 Q_M^I = \int d{\bf S} \cdot \Tr({\bf B}\phi^I)   \; ,
\ee
are the magnetic and electric charge vectors in $\bbbr^6$, respectively,
and $\gamma$ is the angle between them.
Analysis of the Hamiltonian derived from (\ref{action})
reveals a lower bound on the mass of charged states, 
\be
  M^2 \geq Z^2_+ \; .
\ee
This is the Bogomol'nyi-Prasad-Sommerfeld (BPS) bound.
There are two kinds of states (BPS states) that saturate this bound.
For $\gamma = 0$, {\it i.e.} parallel electric and magnetic charge vectors
$Q_M^I \propto Q_E^I$,
the BPS states preserve 1/2 of the underlying supersymmetry,
and therefore transform in a short ($2^4$) representation of the 
superalgebra, in which the highest spin is $1$.
 Their mass is given by
\be
  M^2_{1/2\, BPS} =  |Q^I_E|^2 + |Q^I_M|^2 \; .
\label{1/2BPS}
\ee
For $\gamma > 0$, {\it i.e.} $Q_M^I \propto\!\!\!\!\!\!/ \,\,\,Q_E^I$, the BPS
states preserve 1/4 of the supersymmetry, and therefore
transform in a medium ($2^6$) representation of the superalgebra,
in which the highest spin is $3/2$.
Their mass is given by
\be
 M^2_{1/4\, BPS} =  |Q^I_E|^2 + |Q^I_M|^2 + 
        2|Q_E^I||Q_M^I|\sin{\gamma} \; .
\label{1/4BPS}
\ee
As we shall see, these states can only arise if the rank of the group
is greater than 1.
In both cases the masses are protected from quantum corrections,
since both supermultiplets are shorter than the generic ($2^8$)
representation \cite{ow}.

The moduli space of vacuua is parameterized by the 
VEV's of the adjoint Higgs fields $\phi^I$ in the Cartan subalgebra 
${\bf H}$,
\be
  \langle \phi^I \rangle = {\bf v}^I \cdot {\bf H} \; .
\ee
Let us assume that the gauge group $G$ is maximally broken
to $U(1)^r$, where $r$ is the rank of $G$.
The electric and magnetic charge vectors are then given by
\be 
  Q_E^I = e {\bf q} \cdot {\bf v}^I    \qquad , \qquad 
  Q_M^I = {4\pi\over e} {\bf g} \cdot {\bf v}^I \; ,
\label{charges}
\ee
where ${\bf q}$ and ${\bf g}$ are vectors in the root lattice and co-root 
lattice of $G$, respectively.
As such, they can be expanded in terms of the
simple roots $\bbeta^{(a)}$ and simple co-roots $\bbeta^{(a)*}$,
\be
 \bq = \sum n_e^a\mbox{\boldmath{$\beta$}}^{(a)} \qquad , \qquad
 \bg = \sum n_m^a\mbox{\boldmath{$\beta$}}^{(a)*} \; ,
\ee
where $\bbeta^{(a)*} = \bbeta^{(a)}/\bbeta^{(a)2}$, and 
$a=1,\ldots,r$.

For $G=SU(2)$ the expressions for the charge vectors reduce to
\be
  Q_E^I = e n_e v^I  \qquad , \qquad  Q_M^I = {4\pi\over e} n_m v^I    \; ,
\ee
so we see that $Q_M^I$ and $Q_E^I$ are always parallel,
and therefore that all the BPS states preserve 1/2 of the supersymmetry.
It also follows from (\ref{1/2BPS}) and from the triangular inequality that BPS
states with $(n_e,n_m)$ relatively prime integers are stable for all $e$.
The perturbative BPS spectrum consists of a neutral massless photon
multiplet $(0,0)$, and massive $W^\pm$-boson multiplets $(\pm 1,0)$.
S-duality maps these to BPS states $(n_e,n_m)$ with $n_e$ and $n_m$
relatively prime. 
For $n_m=1$ this is just the 'tHooft-Polyakov $SU(2)$ monopole,
and its dyonic generalizations.
Finding the states with
larger values of $n_m$ is harder, and is equivalent to 
the problem of finding normalizable harmonic forms on the moduli space
of monopoles. This was done for $n_m=2$ in \cite{sen1},
and  substantial evidence was provided for the existence
of such forms for $n_m>2$ as well \cite{ss,porrati}.

The case  $G=SU(3)$ is the first instance where both kinds
of BPS states are allowed.
For $\bn_m\propto\bn_e$, {\it i.e.} $\bg\propto\bq$ and $Q_M^I \propto Q_E^I$, 
the BPS states are in short multiplets (1/2 BPS). 
The massive perturbative BPS states are $W$-bosons 
with $\bn_e = \pm(1,0), \pm(0,1)$ and
$\bn_e = \pm(1,1)$, corresponding
to the two simple roots $\bbeta^{(1)}, \bbeta^{(2)}$, and
the non-simple positive root $\bgamma = \bbeta^{(1)}+\bbeta^{(2)}$. 
S-duality maps these to states 
with $(\bn_e,\bn_m) = (k(1,0),l(1,0)), (k(0,1),l(0,1))$, and $(k(1,1),l(1,1))$ 
respectively, where $k$ and $l$ are relatively prime integers. The 
property of $\bn_m\propto\bn_e$ is preserved, so these are 1/2
BPS states as expected. In fact these are the only BPS states predicted
by S-duality. It follows from the BPS mass formula (\ref{1/2BPS}) 
that the first two classes of states are absolutely stable, whereas
the states obtained from the $\bn_e=(1,1)$ $W$-boson are only marginally
stable.
The former are obtained by embedding $SU(2)$ monopoles with
charge $l$ using the appropriate simple root.
The latter class of states are in general more difficult to find.
The special case of the $\bn_m=(1,1)$ monopole was obtained
by an $SU(2)$ embedding using the root $\bgamma$ \cite{gl}.

Unlike in the $SU(2)$ case, we now have the possibility of choosing
electric and magnetic charge vectors which are not parallel, {\it i.e.} 
$\bn_m\propto\!\!\!\!\!\!/\,\,\,\bn_e$. The corresponding BPS states 
would preserve only 1/4 of the supersymmetry, and would have a mass
given by (\ref{1/4BPS}).
The existence of such states is not predicted by S-duality, since they
cannot be obtained from any perturbative states. In fact they
lie on separate $SL(2,\bbbz)$ orbits.
Furthermore, there are no known solutions to the Bogomol'nyi equations 
with non-parallel electric and magnetic charge vectors. 
Nevertheless, we will show that they do indeed exist.

\subsection{D3-brane construction}

\noindent Four dimensional $N=4$ SYM with a gauge
group $U(n)$ (or $SU(n)$) can be realized
as the low-energy effective theory on the world-volume
of $n$ parallel D3-branes \cite{tseytlin,gg}. As we are primarily interested
in $SU(3)$, let us consider three parallel D3-branes,
extended along the directions $(x^1,x^2,x^3)$.
These correspond to three points $R_1,R_2,R_3$ in $\bbbr^6$,
and therefore define a plane. The generic gauge group
is $U(1)^3$, and is enhanced to $U(3)$ when all three
D3-branes coincide. Since the c.o.m. degrees of freedom
are decoupled from the rest, we shall focus on the $SU(3)$
subgroup of $U(3)$, corresponding to relative degrees of freedom.
This allows us to fix the position of one of the three D3-branes,
so we choose $R_3$ to be at the origin. 
For generic values of $R_1^I$ and $R_2^I$ the gauge
group $SU(3)$ is broken to $U(1)^2$.
We shall work in a basis in which the two
$U(1)$'s are associated to the two D3-branes at $R_1$ and $R_2$,
respectively. In this basis it is easy to identify all the $W$-bosons.
The $\pm(1,0)$ bosons correspond to a fundamental string, of either
orientation, between the D3-brane at $R_3$ and the one at $R_1$;
the $\pm(0,1)$ bosons correspond to a fundamental string between
$R_3$ and $R_2$, and the $\pm(1,1)$ bosons -- to a fundamental string
between $R_1$ and $R_2$ (Fig.~1).

\begin{figure}[htb]
\epsfysize=4cm
\centerline{\epsffile{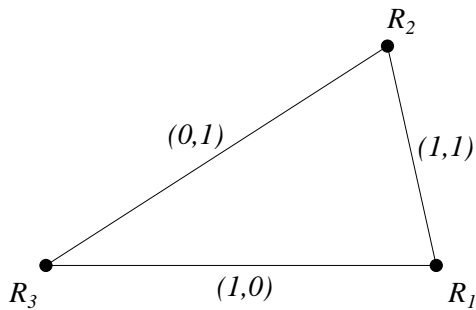}}
\caption{D3-brane representation of $N=4$ $SU(3)$
SYM.}
\end{figure}

It is also straightforward to identify all the monopole and dyon states
predicted by S-duality. These correspond to D-strings, and bound
states of D-strings and fundamental strings, respectively. These
bound states
are predicted by S-duality in type IIB string theory, and were shown
to exist in \cite{wittenboundstates}.

The D-brane separations are proportional to the Higgs VEV's
in the SYM theory.
We can determine the precise numerical factors by comparing
the masses of $W$-bosons and monopoles in the two
approaches. The result can be written as\footnote{Masses are 
measured in the Einstein metric.}
\be
  R_a^In_e^a = {2\pi\alpha'\over \sqrt{g_s}} Q_E^I \quad , \quad
  R_a^In_m^a = {2\pi\alpha' \sqrt{g_s}} Q_M^I \; ,
\label{QtoR}
\ee
where $g_s$ is the type IIB string coupling, and 
with $Q_E^I$ and $Q_M^I$ given in (\ref{charges}).

\section{Three-pronged strings and D-branes}
\setcounter{equation}{0}

\noindent Type IIB string theory contains, in addition to an
$SL(2,\bbbz)$ multiplet of $(p,q)$ strings, objects consisting
of three strings meeting at a point (Fig.~2). These are known as
``three-string junctions" \cite{schwarz}, or 
``three-pronged strings"
\cite{gz}. If the three strings are of type $(p_i,q_i)$ charge conservation 
requires 
\be
 \sum_{i=1}^3 p_i = \sum_{i=1}^3 q_i = 0 \; .
\label{stringcharges}
\ee
We shall therefore refer to these as 
``$((p_1,p_2),(q_1,q_2))$ strings".

\begin{figure}[htb]
\epsfysize=4cm
\centerline{\epsffile{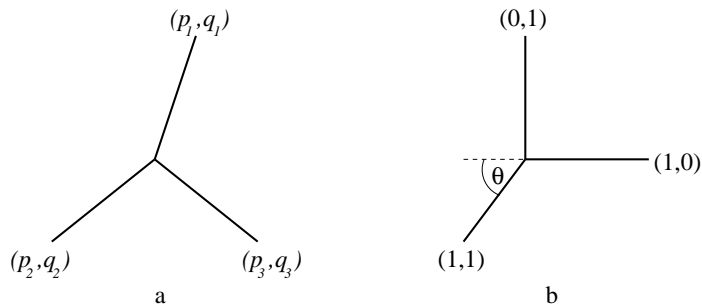}}
\caption{Three-pronged string, (a) in the general case,
and (b) for $(1,0),(0,1)$ and $(1,1)$ prongs.
The angle between the $(1,0)$ and $(0,1)$ prongs
is $90^\circ$, and the angle $\theta$ is given by
$\tan\theta = 1/g_s$.}
\end{figure}
The angles between the strings are determined by the requirement
that the net force on the junction point vanishes \cite{schwarz}.
This is achieved if
\be 
 \sum_{i=1}^3 T_{p_i,q_i} \widehat{n}_i = 0 \; ,
\label{zeroforce}
\ee
where $\widehat{n}_i$ is the direction of the $i$'th string and 
$T_{p_i,q_i}$ is its tension. Recall that the tension of a $(p,q)$ string
is given in the Einstein metric by
\be
 T_{p,q} = {\sqrt{g_s}\over 2\pi\alpha'} | p + q\tau| \; ,
\label{stringtension}
\ee
where $\tau = i/g_s + \chi/2\pi$, and $\chi$ is the expectation 
value of the type IIB R-R scalar field. We shall assume 
that $\chi=0$, which is consistent with our assumption that 
$\theta_{YM}= 0$.\footnote{Generalizations for $\chi\neq 0$, and
thus $\theta_{YM}\neq 0$, will be discussed elsewhere.}

It follows from (\ref{stringcharges}), ({\ref{zeroforce}), and (\ref{stringtension})
that a $(p,q)$ string which is part of a three-string junction, or more generally
a string network, must be oriented along the vector $(p + q\tau)$ in the 
(complex) plane of the network \cite{sen2}. 
This in turn implies that 
the unbroken supersymmetries of {\em any} three-pronged string,
or string network, are given by $\epsilon_L Q_L + \epsilon_R Q_R$, with \cite{sen2}
\begin{eqnarray}
  \epsilon_L &=& \Gamma_1\cdots\Gamma_8\epsilon_L \nonumber \\
  \epsilon_R &=& - \Gamma_1\cdots\Gamma_8\epsilon_R \nonumber \\
  \epsilon_L &=& \Gamma_1\cdots\Gamma_7\Gamma_9\epsilon_R \; ,
\label{susy1}
\end{eqnarray}
where we have assumed that the network lies in the $(x^8,x^9)$-plane.
These leave 1/4 of the original type IIB supersymmetry unbroken.
The first two conditions give the unbroken supersymmetries of
a fundamental string extended along $x^9$.
The last condition can be turned into something more familiar using
the ten-dimensional chirality operator. The original 16-component 
spinors have a definite ten-dimensional chirality,
\be
  \Gamma^{11}\epsilon_L = + \epsilon_L \qquad , \qquad 
  \Gamma^{11}\epsilon_R = + \epsilon_R \; ,
\ee
where $\Gamma^{11} = \Gamma_0\cdots\Gamma_9$. This means that
\be 
 \Gamma_0\Gamma_9\epsilon_L = \Gamma_1\cdots\Gamma_8\epsilon_L
   = \epsilon_L \; ,
\ee
where the last equality is a result of the first condition in (\ref{susy1}).
Multiplying the third condition in (\ref{susy1}) by $\Gamma_0\Gamma_9$ 
then gives 
\be
  \epsilon_L = \Gamma_0\cdots\Gamma_7\epsilon_R \; ,
\label{d7brane}
\ee
which, by itself, gives the unbroken supersymmetries of a D7-brane
extended along the directions $(x^1,\ldots, x^7)$.
In summary, the supersymmetry preserved by a three-pronged string
is the same as the supersymmetry preserved by a 
fundamental string oriented in the plane defined by the three prongs, together
with a D7-brane transverse to that plane.

Since we know normal open strings can end on D-branes, whilst preserving 
a certain amount of supersymmetry, it is natural to ask whether the
same is true for three-pronged strings. Since the  supersymmetry
conditions of the three-pronged strings already include effectively 
a D7-brane transverse to the plane of the three-pronged string, we
conclude that adding D7-branes transverse to that plane  
does not break any more supersymmetry. On the other hand
adding D3-branes
or D(-1)-branes transverse to the plane further breaks the supersymmetry
by half, leaving 1/8 of the original amount, and adding transverse
D1-branes, D5-branes or D9-branes breaks all the supersymmetry.

Since the three prongs are mutually non-local, they cannot all
end on D-branes in general. The exception is the D3-brane,
on which any $(p,q)$ string can end. 
Three parallel D3-branes can therefore be connected
by a three-pronged string, lying in a plane transverse
to the D3-branes (Fig.~3). Furthermore, this 
configuration preserves
1/8 of the original supersymmetry.
Of course since we do not yet know how to quantize
three-pronged strings, our argument relies on the assumption 
that 
$(p,q)$ prongs are equivalent to $(p,q)$ strings as far
as boundary conditions go.

\begin{figure}[htb]
\epsfysize=4cm
\centerline{\epsffile{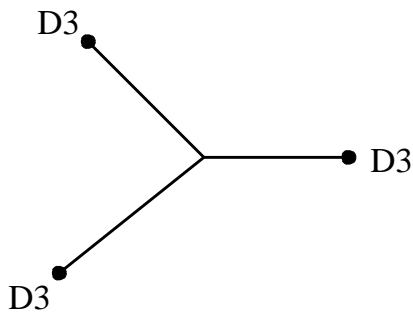}}
\caption{Three-pronged string ending on three D3-branes,
which are transverse to the plane defined by the sting.}
\end{figure}

The analogous situation with 7-branes is less clear,
as those would have to be mutually non-local. It is not
clear how much, if any, supersymmetry is left unbroken
by a generic system of parallel, mutually non-local, 7-branes.
Certain configurations of mutually non-local 7-branes, namely
those that arise from F-theory compactification on K3, are known
to preserve 1/2 of the supersymmetry. In those situations one could
connect the 7-branes with the appropriate three-pronged strings,
leaving 1/4 of the supersymmetry unbroken. Such configurations
were studied in \cite{gz}, in the context of exceptional gauge
enhancement in eight dimensions.

In what follows we shall focus on the D3-brane configuration of section~2,
in which the three D3-branes are located at the points $0$, $R_1$, and $R_2$
in the $(x^8,x^9)$ plane.

\section{1/4 BPS States and Marginal Stability}
\setcounter{equation}{0}
\subsection{1/4 BPS states}
\noindent
Since the configuration of three parallel D3-branes connected
by a three-pronged string preserves 1/8 of the space-time 
supersymmetry, namely 4 supersymmetries, it also preserves
1/4 of the D3-brane world-volume supersymmetry.
This strongly suggests that we identify the corresponding
world-volume states with the 1/4 BPS states of $N=4$ SYM discussed in
section~2. 
In fact, since the three prongs, and in particular the
two prongs ending on the D3-branes at $R_1^I$ and $R_2^I$,
are mutually non-local, the corresponding electric and magnetic 
charge vectors $\bn_e,\bn_m$ will be non-parallel, as required
for the 1/4 BPS states.
Our conjecture is then that 
the ground state of the $((p_1,p_2),(q_1,q_2))$ string corresponds in
the world-volume theory to the 1/4 BPS state with 
$(\bn_e,\bn_m) = ((p_1,p_2),(q_1,q_2))$.

As further evidence for this conjecture, let us compute the mass
of a specific 1/4 BPS state, and compare it with the mass of the
corresponding three-pronged string.
With the aid of (\ref{QtoR}) we can rewrite the mass formula for
1/4 BPS states (\ref{1/4BPS}) in terms of the geometric variables $R_a^I$ as
\be
  M^2_{1/4\, BPS} = \left({1\over 2\pi\alpha'}\right)^2
     \left(g_s|R_a^In_e^a|^2 + {1\over g_s}|R_a^In_m^a|^2
     + 2 |R_a^In_e^a| |R_a^In_m^a| \sin\gamma\right) \; .
\label{1/4BPS2}
\ee
Let us now consider a specific state, with
\be
 (\bn_e,\bn_m) = ((1,0),(0,1)) \; .
\ee
This state is electrically charged under the first $U(1)$,
and magnetically charged under the second $U(1)$. Its 
mass is given by
\be 
  M^2_{((1,0),(0,1))} = \left({1\over 2\pi\alpha'}\right)^2
     \left(g_s|R_1|^2 + {1\over g_s}|R_2|^2
     + 2 |R_1| |R_2| \sin\gamma\right) \; .
\label{mass1}
\ee
The corresponding three-pronged string has charges 
\be
 (p_1,q_1) = (1,0) \quad , \quad (p_2,q_2) = (0,1) \quad , \quad 
 (p_3,q_3) = (-1,-1) \; .
\label{basic}
\ee
The orientations of the prongs are shown in 
Fig.~2b and Fig.~4.
The $(1,0)$ and $(0,1)$ prongs meet at a right angle, and the 
$(1,1)$
prong comes in at an angle $\theta$ given by
\be 
  \tan\theta = 1/g_s \; .
\label{theta}
\ee
Let us assume that all the angles of the triangle formed 
by the three D3-branes $\alpha,\beta,\gamma$ are less than 
$90^\circ$. 
We shall discuss
what happens when one of these angles
grows beyond $90^\circ$ in the next subsection.

\begin{figure}[htb]
\epsfysize=5cm
\centerline{\epsffile{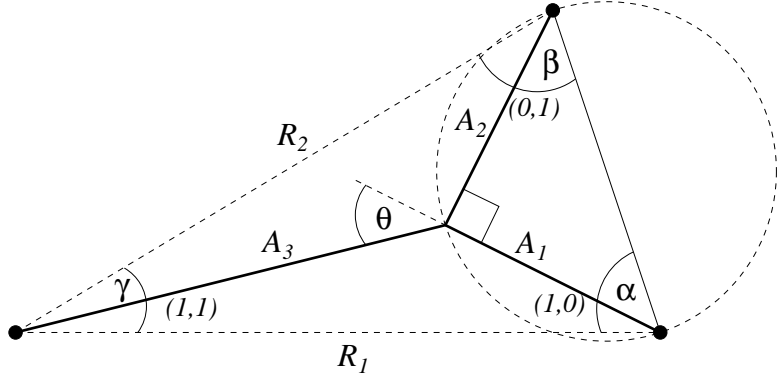}}
\caption{A $((1,0),(0,1))$ three-pronged string ending on
three D3-branes at $R_1^I,R_2^I$ and $0$. The angles
of the triangle defined by these points are
$\alpha,\beta$ and $\gamma$, respectively, and the lengths
of the corresponding prongs are $A_1,A_2$ and $A_3$.
The angle $\theta$ is given by $\tan\theta = 1/g_s$.}
\end{figure}

The mass of the three-pronged string connecting the three D3-branes
is simply given by the sum of the masses of the individual prongs:
\begin{eqnarray}
 M^2_{3-string} &=& \Big(T_{(1,0)}A_1 + T_{(0,1)}A_2 
      + T_{(-1,-1)}A_3\Big)^2\nonumber \\
  &=& {g_s\over (2\pi\alpha^\prime)^2}
            \left( A_1 + {1\over g_s}A_2 + \sqrt{1+{1\over g_s^2}}A_3 
\right)^2 \; .
\label{mass2}
\end{eqnarray}
From the geometry of Fig.~4 it follows that
the lengths of the prongs $A_1,A_2,A_3$ are
related to the lengths $R_1,R_2$ and the angle $\gamma$ by
\begin{eqnarray}
 R_1^2 &=& A_1^2 + A_3^2 + 2A_1A_3\cos{\theta} \nonumber \\
 R_2^2 &=& A_2^2 + A_3^2 + 2A_2A_3\sin{\theta} \nonumber \\
 A_1^2 + A_2^2 &=& R_1^2 + R_2^2 - 2|R_1||R_2|\cos{\gamma} \; .
\end{eqnarray}
These can be combined to get a fourth relation,
\be
 |R_1||R_2|\sin\gamma = A_1A_2 + A_1A_3\sin\theta + A_2A_3\cos\theta \; .
\ee
Using (\ref{theta}) together with the above relations in the mass formula
(\ref{mass2}), we find
\be
 M^2_{3-string}  = \left({1\over 2\pi\alpha'}\right)^2
     \left(g_s|R_1|^2 + {1\over g_s}|R_2|^2
     + 2 |R_1| |R_2| \sin\gamma\right) \; .
\label{3string}
\ee
This agrees with the mass of the $((1,0),(0,1))$ BPS state
(\ref{mass1}).
We therefore conclude that the ground state of the $((1,0),(0,1))$ three-pronged
string does indeed correspond to the $((1,0),(0,1))$ 1/4 BPS state in
$N=4$ SYM.
Other 1/4 BPS states can be obtained from this one by the action of 
the field-theoretic duality group $SL(2,\bbbz)$.
On the other hand other three-pronged strings can be obtained from the 
basic one (\ref{basic}) by the action of the type IIB duality group
$SL(2,\bbbz)$. Since in both cases the states transform as doublets
under $SL(2,\bbbz)$, we conclude that the correspondence between
three-pronged strings and 1/4 BPS states extends to all charges that 
can be obtained from the ones above by S-duality. 

When the three D3-branes coincide, {\it i.e.} when $R_1=R_2=0$,
these states become massless. The medium multiplets then decompose
into short multiplets, some of which contain particles with spin $>1$.
The existence of massless particles with spin $>1$ seems surprising
at first sight, but is actually expected since the theory is
conformal.

\subsection{marginal stability}

\noindent 
It follows from (\ref{1/4BPS2}) that 
\be
  M_{((1,0),(0,1))} \leq M_{((1,0),(0,0))} + M_{((0,0),(0,1))} \; ,
\ee
where the inequality is saturated for $\gamma=90^\circ$. This suggests that
when $\gamma=90^\circ$, the above 1/4 BPS state is only marginally stable, and may
decay into the purely electric state $((1,0),(0,0))$ and the purely magnetic state
$((0,0),(0,1))$. It does not however mean that the state does indeed decay.
In particular the inequality still holds for $\gamma > 90^\circ$, so it appears
that if the above BPS state still exists it should be stable.

On the other hand it is clear from the three-pronged string representation
of this state that it does not exist if $\gamma > 90^\circ$, since the $(1,0)$
and $(0,1)$ prongs would have to meet outside of the triangle formed by 
the three D3-branes (Fig.~5c). 
This representation also provides the mechanism by which the
$((1,0),(0,1))$ state decays. For $\gamma < 90^\circ$ (Fig.~5a) 
there exists a 
three-pronged string, with the required charges, connecting 
the D3-branes. It follows from (\ref{3string}) that this state is
lighter than the state consisting of separate $(1,0)$ and $(0,1)$ strings.
As $\gamma$ is increased, the length of the $(1,1)$
prong decreases, until it vanishes when $\gamma = 90^\circ$
(Fig.~5b). At this point
we just have a $(1,0)$ string and $(0,1)$ string that meet at a point on the
common D3-brane. There is now a flat direction associated to
separating the two strings along this D3-brane. This
corresponds in the world-volume theory to the decay of the 1/4
BPS state to a purely electric 1/2 BPS state and a purely magnetic 
1/2 BPS state. For $\gamma > 90^\circ$ (Fig.~5c) the two string state is the only
one carrying the above charges, so we conclude that $\gamma = 90^\circ$
defines a curve of marginal stability $\C_3$ on the plane of
the three-pronged string, given by the smallest circle passing through
$R_1$ and $R_2$.

\begin{figure}[htb]
\epsfysize=4cm
\centerline{\epsffile{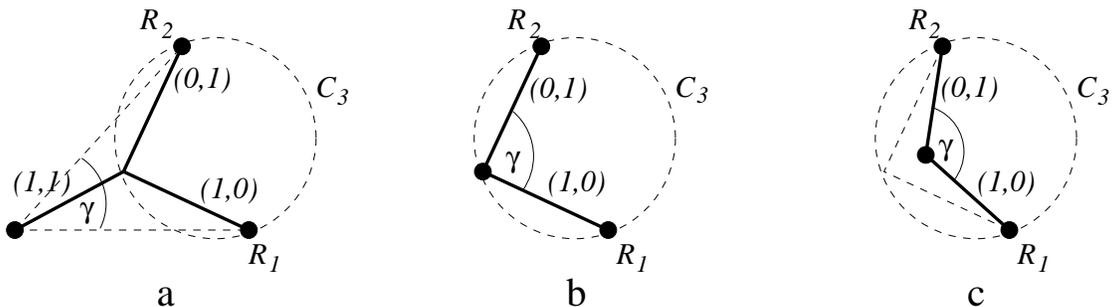}}
\caption{Moving a D3-brane through the curve of marginal
stability. In (a) $\gamma<90^\circ$, and the D3-branes
can be connected by the $((1,0),(0,1))$ three-pronged
string. In (b) $\gamma=90^\circ$, and the three-pronged
string becomes degenerate with a $(1,0)$ string plus
a $(0,1)$ string. In (c) $\gamma<90^\circ$, and only
the two open string state exists.}
\end{figure}

More generally we can see from the geometry of Fig.~4 that the
conditions on the angles for a $((1,0),(0,1))$ three-pronged string  
to exist are
\be
 \alpha < \theta + 90^\circ \qquad , \qquad
 \beta < 180^\circ - \theta \qquad , \qquad
 \gamma < 90^\circ \; ,
\ee
where $\theta$ is determined by the value of the string coupling constant
(\ref{theta}).
We have argued that when $\gamma$ exceeds its bound the three-pronged
string decays into a $(1,0)$ string and a $(0,1)$ string. 
On the other hand, if $\alpha$ exceeds its bound the three-pronged
string will decay into a $(0,1)$ string and a $(1,1)$ string.
In the world-volume theory these correspond to a monopole
of charge $(\bn_e,\bn_m)=((0,0),(-1,1))$, and a dyon of charge 
$(\bn_e,\bn_m) = ((1,0),(1,0))$, respectively. Similarly if $\beta$ exceeds its bound
the three-pronged string will decay into a $(1,0)$ string and a $(1,1)$
string, which correspond to a $W$-boson of charge
$(\bn_e,\bn_m)=((1,-1),(0,0))$  and a dyon of charge
$(\bn_e,\bn_m)=((0,1),(0,1))$.

The conditions on $\alpha$ and $\beta$ also define curves of marginal stability
$\C_1,\C_2$ passing through the the D3-branes at $0,R_2^I$ and $0,R_1^I$,
respectively. 
The curve $\C_1$, for example, is
given by the set of points $\{Q\}$ in the $(0,R_1^I,R_2^I)$ plane
for which the angle between the segments $Q0$ and $QR_2$ is $\theta + 90^\circ$.
The curve $\C_2$ is defined similarly.

\section{Summary and Outlook}

\noindent
We have shown that a configuration of three parallel D3-branes connected
by a three-pronged string preserves 1/8 of original space-time 
supersymmetry, and therefore 1/4 of the D3-brane world-volume
supersymmetry. This, together with the fact that the world-volume
electric and magnetic charge vectors were not parallel,
was used to argue that the ground state
of such a string corresponds to a 1/4 BPS state in the world-volume
$N=4$ SYM theory. As further evidence for this conjecture we showed
that the masses agree.

The D3-brane approach has allowed us to identify curves of marginal 
stability in the moduli space of $N=4$ SYM. On these curves the 1/4 
BPS states are marginally stable to decay into two 1/2 BPS states.
In the D3-brane picture this is seen as the dissociation of the 
three-pronged
string into two open strings when one of the prongs degenerates.

Strictly speaking, our result applies to the 1/4 BPS states
which can be obtained from the $(\bn_e,\bn_m)=((1,0),(0,1))$ state
by $SL(2,\bbbz)$, and for $\theta_{YM} = 0$. It would be interesting
to generalize it for $\theta_{YM} \neq 0$. But more importantly
we would like to generalize the result to other $SL(2,\bbbz)$
conjugacy classes. In particular, the representation of such
states as three-pronged strings may simplify the problem of counting
them, and this may be relevant to black-hole entropy.

Three-pronged strings may also play a role in four-dimensional $N=2$
superconformal field theories, which were studied in \cite{ad,apsw}.
From the point of view of type IIB string theory these correspond
to the world-volume theories of a D3-brane, when it coincides
with two mutually non-local 7-branes. 

There are still many unanswered questions, the most important
of which is how to actually quantize three-pronged strings, and thus
get a better handle on their space-time properties.

\section*{Acknowledgments}

I would like to thank Ansar Fayyazuddin, Cumrun Vafa, and
Barton Zwiebach for useful discussions. I would also like to 
thank Matthias Gaberdiel for a critical reading of
the manuscript.
This work is supported in part by the NSF under grant 
PHY-92-18167.

\end{document}